\documentclass[fleqn,10pt]{wlscirep}
\usepackage[utf8]{inputenc}
\usepackage[T1]{fontenc}
\usepackage{graphicx}% Include figure files
\usepackage{color}
\usepackage{dcolumn}% Align table columns on decimal point
\usepackage{latexsym}
\usepackage[normalem]{ulem}
\usepackage{hyperref,amssymb}
\usepackage{url}
\usepackage{color,soul}
\usepackage{graphicx}
\usepackage{verbatim}
\usepackage{multirow}
\usepackage{amsmath}
\newcommand{\beq}{\begin{eqnarray}}
\newcommand{\eeq}{\end{eqnarray}}
\usepackage{mathrsfs}
\usepackage{float,soul}
\usepackage{mathtools}
\usepackage{slashed}
\usepackage{physics}	% installed packages for typing maths and physics
\usepackage{graphicx}   % need for Fig.ures
\usepackage{epstopdf}
\usepackage{tabularx}
\usepackage{subfigure}  % use for side-by-side Fig.ures
\usepackage{hyperref}   % use for hypertext links, including those to external documents and URLs
\usepackage{bbold}
\usepackage{wasysym}
\usepackage{feynmp}
\usepackage{hyperref}
\hypersetup{colorlinks,
}

\usepackage{siunitx}
\usepackage{soul}
\usepackage{relsize}
\soulregister{\cite}7
\soulregister{\ref}7
\soulregister{\eqref}7
\usepackage{amsmath}

\usepackage{booktabs}

\DeclareSIUnit\angstrom{\text{\AA}}
\title{On the temperature dependence of the density of states of liquids at low energies}

\author[1,2,3+]{Sha Jin}
\author[4,5,6,+]{Xue Fan}
\author[7,8]{Caleb Stamper}
\author[8]{Richard A. Mole}
\author[1]{Yuanxi Yu}
\author[1,4,9]{Liang Hong}
\author[8,*]{Dehong Yu}
\author[1,2,3,*]{Matteo Baggioli}
\affil[1]{School of Physics and Astronomy, Shanghai Jiao Tong University, Shanghai 200240, China}
\affil[2]{Wilczek Quantum Center, Shanghai Jiao Tong University, Shanghai 200240, China}
\affil[3]{Shanghai Research Center for Quantum Sciences, Shanghai 201315, China}
\affil[4]{Shanghai National Center for Applied Mathematics, Shanghai Jiao Tong University, Shanghai 200240, China}
\affil[5]{Materials Genome Institute, Shanghai University, Shanghai 200444, China}
\affil[6]{School of Materials Science and Engineering, Georgia Institute of Technology, Atlanta, GA 30332, United States of America}
\affil[7]{Institute for Superconducting and Electronic Materials, University of Wollongong, Wollongong, NSW 2500, Australia}
\affil[8]{The Australian Nuclear Science and Technology Organisation, Lucas Heights, NSW 2232, Australia}
\affil[9]{Institute of Natural Sciences,Shanghai Jiao Tong University, Shanghai 200240, China}

\affil[*]{Corresponding authors: dyu@ansto.gov.au, b.matteo@sjtu.edu.cn}

\affil[+]{These authors contributed equally to this work}

\begin{abstract}
We report neutron-scattering measurements of the density of states (DOS) of water and liquid Fomblin in a wide range of temperatures. In the liquid phase, we confirm the presence of a universal low-energy linear scaling of the experimental DOS as a function of the frequency, $g(\omega)= a(T) \omega$, which persists at all temperatures. The low-frequency scaling of the DOS exhibits a sharp jump at the melting point of water, below which the standard Debye's law, $g(\omega) \propto \omega^2$, is recovered. On the contrary, in Fomblin, we observe a continuous transition between the two exponents reflecting its glassy dynamics, which is confirmed by structure measurements. More importantly, in both systems, we find that the slope $a(T)$ grows with temperature following an exponential Arrhenius-like form, $a(T) \propto \exp(-\langle E \rangle /T)$. We confirm this experimental trend using molecular dynamics simulations and show that the prediction of instantaneous normal mode (INM) theory for $a(T)$ is in qualitative agreement with the experimental data. 
\end{abstract}
\begin{document}

\flushbottom
\maketitle

\thispagestyle{empty}

\section*{Introduction}
The density of states (DOS) is a fundamental concept in solid state physics, and plays a key role in determining vibrational, thermodynamic and transport properties of a given material (\textit{e.g.}, heat capacity, thermal conductivity, superconductivity, etc.). In crystalline solids with long-range order, the definition and determination of the DOS can be achieved using a normal mode analysis, which ultimately leads to the construction of the Debye model \cite{kittel2021introduction}. The Debye model predicts that the DOS of 3D crystalline solids, in the low frequency regime, displays a universal quadratic scaling known as Debye's law,
\begin{equation}\label{ee1}
    g_{\text{crystals}}(\omega) \propto \frac{\omega^2}{\bar{v}^3},\qquad \text{where} \qquad  \frac{1}{\bar{v}^3}=\frac{2}{v_T^3}+\frac{1}{v_L^3},
\end{equation}
with $v_{T,L}$ respectively the speed of propagation of transverse and longitudinal phonons.
The Debye model has been very successful in describing the dynamics of crystalline solids \cite{kittel2021introduction}.

Despite a lot of effort having been dedicated to investigate and explain deviations from Debye's law in the DOS of amorphous systems \cite{doi:10.1142/q0371}, much less is known for the case of classical liquids. In particular, the properties of the DOS of liquids at low energies, in the range where Debye's law holds in crystalline solids, remains unclear.

The experimental investigation of the density of states of liquids at low energies is much less explored due to the fact that many techniques used to study liquid properties, such as infrared, Raman and nonlinear IR, do not have sufficient energy resolution to probe the details at very low energies. To the best of our knowledge, the first experimental investigation in this direction was presented by Phillips et al. in 1989 \cite{PhysRevLett.63.2381}. A comparison of the low-frequency experimental density of states on an absolute scale for glassy, liquid, and polycrystalline selenium suggests that the DOS of liquids does not obey Debye's law.
Eleven years later, Dawidowski et al.\cite{DAWIDOWSKI2000247} performed an experimental study of the density of states of heavy water by comparing two temperatures slightly above and slightly below the freezing transition. Once again, their experimental data indicate the disappearance of Debye scaling in the liquid phase.

Along with the development of modern cold neutron spectrometers and intense inelastic X-ray spectrometers based on synchrotron radiation facilities, this field has become more active. Recent experimental results of the density of states, using inelastic neutron scattering (INS), in several liquid systems including water, liquid metal and polymer liquids have been reported in Ref. \cite{doi:10.1021/acs.jpclett.2c00297} (see also Ref.\cite{doi:10.1021/jp0641132} and Ref.\cite{doi:10.1146/annurev.physchem.50.1.571} for previous studies using neutrons). These more recent results confirmed the existence of a universal linear in frequency law in the DOS of liquids, $g(\omega)\propto \omega$, compatible with the experimental data of \cite{PhysRevLett.63.2381,DAWIDOWSKI2000247}.

To some degree, there is experimental evidence that the low-energy density of states of liquids follows a universal linear scaling in frequency of the form:
\begin{equation}
    g_{\text{liquids}}(\omega) = a(T) \omega, \label{figdue}
\end{equation}
where $a(T)$ is an unkown function of the temperature $T$ which we will refer to in the rest of this manuscript as the ``slope''. Nevertheless, \begin{itemize}
    \item[(I)] It is not clear how universal the form in Eq.\eqref{figdue} is, in which range of energies it emerges, and if it holds at any temperature;
    \item[(II)] It is not known how this linear scaling transitions into the more standard Debye law, Eq.\eqref{ee1}, by approaching the freezing transition;
    \item[(III)] It remains unclear how the slope $a(T)$ depends on temperature and whether the resulting functional dependence could be rationalized with any of the existing theoretical paradigms.
\end{itemize}

The main objective of this work is to answer these three questions.\\

In order to achieve this task, we perform a full scan of the experimental DOS of water and Fomblin oil in a wide range of temperatures, below and above the melting transition, and with particular emphasis in the liquid phase. Additionally, we combine molecular dynamics simulations with theoretical methods to compare the predictions of instantaneous normal mode theory of liquids with our experimental results.

\section*{Experimental results}
We performed INS measurements on two liquids with drastically different elemental compositions, molecular weights, and, therefore, dynamic properties. Water, ``the universal solvent”, is undoubtedly the most well-studied chemical system and so makes a great reference. Nonetheless, water is a scientifically interesting system with complex dynamics that lead to unique chemical properties. To compare with water, we chose to measure Fomblin, a perfluorinated polyether fluid most commonly used as a lubricant for vacuum pumps \cite{test2}, but also technologically relevant for cold neutron storage \cite{pokotilovski2008experimental}. Furthermore, Fomblin is also identified as a low-background pressure-medium for high-pressure neutron scattering experiments \cite{refId0}. In this work, we measured Fomblin-Y, 25/6, with the chemical formula CF$_3$O[-CF(CF$_3$)CF$_2$O-]$_x$(-CF$_2$O-)$_y$CF$_3$ where x/y gives an average molecular weight of $3,300$. The thermodynamic properties of both systems are technologically important and distinct from one another. A summary of some chemical properties relevant to their dynamics is given in Table \ref{tabo}.

\begin{table}[h!]
\centering
\begin{tabular}{ccc}
  \textbf{Property}&\textbf{Water} & \textbf{Fomblin}\\
  \hline 
 Molecular weight (amu) & 18 & 3,300 (av.) \cite{test2}\\
 Density (g cm$^{-3}$) & 1.00 \cite{alma9916097483406676} & 1.90 \cite{test2}\\
 Specific heat (J Kg$^{-1}$ K$^{-1}$) & 4130 \cite{10.1063/1.1461829} & 1000 \cite{test2}\\
 Kinematic viscosity (mm$^2$ s$^{-1}$) & 1 \cite{van1990water}& 276 \cite{test2}\\
 Surface tension (mN m$^{-1}$) & 72 \cite{van1990water} & 22 \cite{test2}\\
\end{tabular}
\caption{Summary of room-temperature chemical properties of water and Fomblin. ``av.'' stands for average.}\label{tabo}
\end{table}

\begin{figure}[h]\centering
\includegraphics[width=\textwidth]{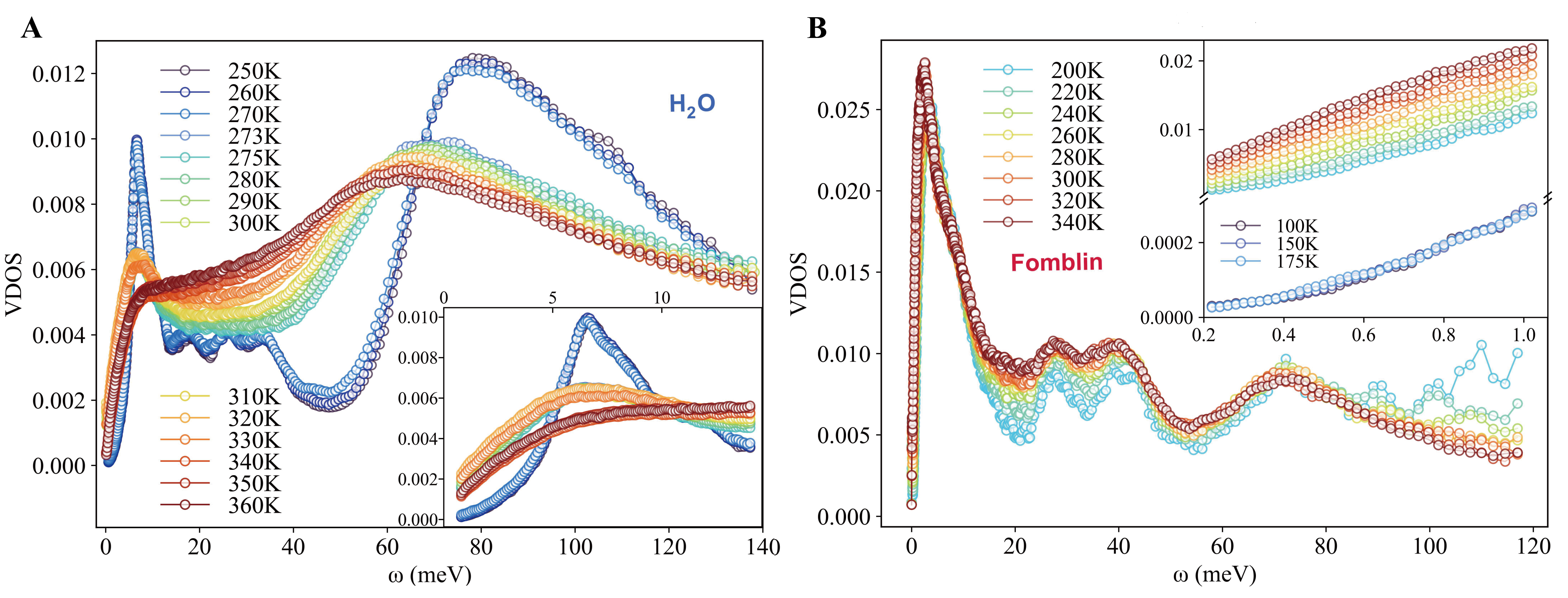}
\caption{The experimental density of states (DOS), measured by INS for different temperatures for \textbf{(A)} water \textbf{(B)} and Fomblin oil. The DOS curves have been normalized by the total area. The insets zoom on the low frequency region to contrast the linear scaling $g(\omega) \propto \omega$ with the Debye scaling $g(\omega) \propto \omega^2$.}
\label{fig:figure1}
\end{figure}

In Fig.\ref{fig:figure1}A, we present the experimental DOS for water as a function of temperature from $250$ K to $360$ K measured with inelastic neutron scattering (see method for DOS derivation). Two major peaks are observed in the DOS spectrum. The first peak appears around $\approx 6.5$ meV and corresponds to the hydrogen-bond bending, perpendicular to the hydrogen-bond (O-H--O) \cite{franks2013water}. The large peak at around $65$ meV is attributed to librational motion due to intermolecular coupling \cite{toukan1988neutron}. In the region between $20-35$ meV, the DOS is characterized by a flat band which represents the weak hydrogen-bond stretching modes in line with the hydrogen-bond \cite{amann2016x}. 

As shown in the inset of Fig.\ref{fig:figure1}A, the DOS exhibits a universal linear behavior at low frequencies in the liquid phase, with temperature above $273$ K. This observation confirms the presence of a low-frequency linear scaling regime, $g(\omega)\propto \omega$, in the whole liquid phase, independently of the temperature, and it expands the results of Ref.\cite{doi:10.1021/acs.jpclett.2c00297}. Let us notice that this linear behavior is not expected to extend all the way down to zero frequency, since the zero frequency value of the DOS is finite due to the diffusive processes active in the liquid phase \cite{hansen2013theory}. 

Below the melting temperature, in the solid phase, the low-energy DOS is drastically modified. First, the zero frequency value disappears, as a confirmation that the self-diffusion constant vanishes in the solid phase. Second, the low-frequency scaling is suddenly modified to recover the standard quadratic Debye law, $g(\omega) \propto \omega^2$. For temperatures below $273$ K, the system is in the solid ice phase and the linewidth of the peak corresponding to the hydrogen-bond bending is significantly smaller. Moreover, the amplitude of the two major peaks is larger than in the liquid phase. As shown in the inset of Fig.\ref{fig:figure1}A, upon the phase transition from solid to liquid, the well-defined peak around $6.5$ meV becomes significantly broadened with much high-intensity signal shifted to low energies. Around $330$ K, in the liquid phase, the lowest peak $\approx 6.5$ meV becomes completely overdamped and the DOS becomes flat up to the next broadened and red-shifted peak at $\approx 65$ meV. This tells us that the dynamics become very unstable with a wide distribution in frequency.

\begin{figure}[h]\centering
\includegraphics[width=0.97\linewidth]{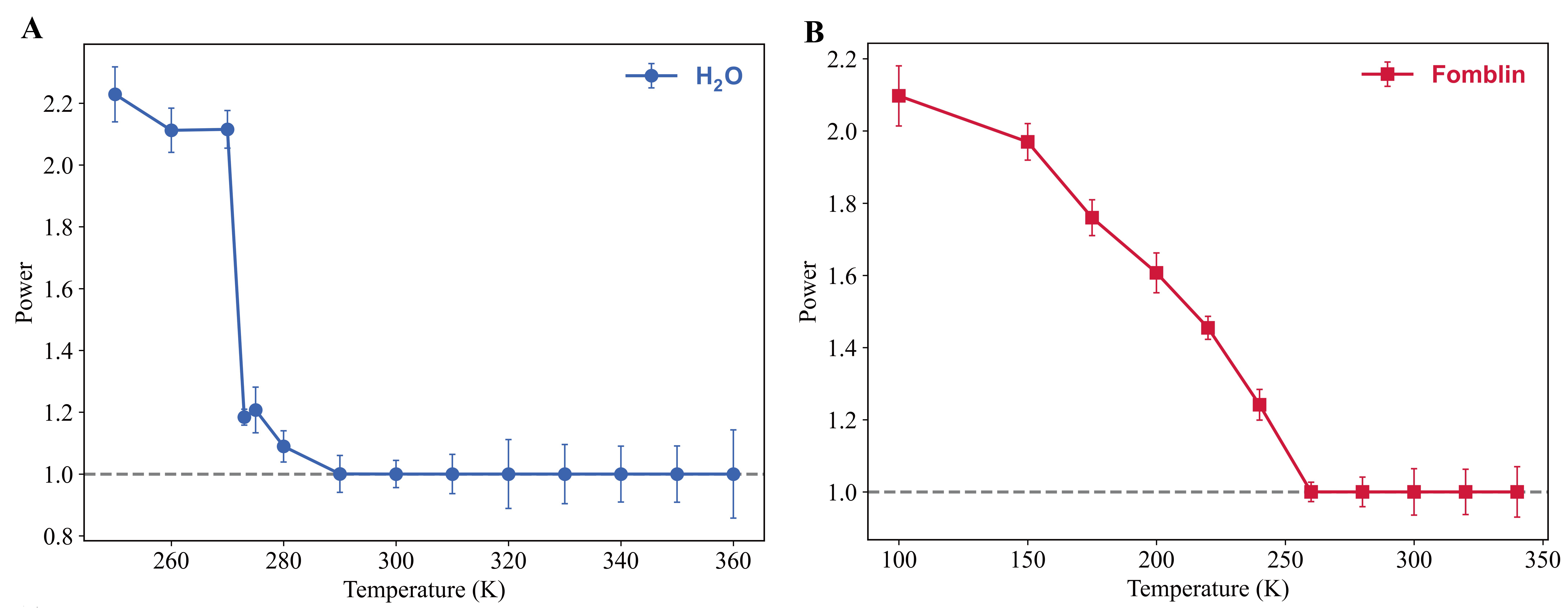}

\caption{The power-law $b(T)$ of the low-frequency experimental DOS as a function of the temperature for \textbf{(A)} liquid water and \textbf{(B)} Fomblin oil. The data are extracted by fitting the experimental DOS with Eq.\eqref{fitfit}. The horizontal gray dashed lines indicate the linear power, $b(T)=1$.}
\label{fig:figure3}
\end{figure}

In Fig.\ref{fig:figure1}B, we show the DOS measured by INS for the Fomblin oil. The lowest band in the DOS has been suggested to correspond to the low-frequency modes caused by the torsion of fluoromethyl ($O-CF_3$) groups at the end of the chain, and C-C torsions of the chain. In the range of  $25-45$ meV, the deformational in-plane modes ($\delta [O-C-C], \delta [O-C-F] $ and $\delta [C-O-C]$) represent the dominant contribution to the DOS spectra \cite{pokotilovski2008experimental}. As shown in the inset of Fig.\ref{fig:figure1}B, and analyzed in more detail in Fig.\ref{fig:figure3}B later, the transition between a low-frequency quadratic scaling and a linear law in Fomblin is more gradual, revealing the absence of a proper first-order melting transition. Nonetheless, also for Fomblin the DOS is linear in the low energy region for all temperatures above $260$ K.

The experimental data for water and Fomblin presented in Fig.\ref{fig:figure1} are taken at different temperatures, and go beyond the solid-liquid phase transition for the two systems. In order to understand how the universal linear scaling characteristic of the liquid phase is modified as a function of temperature entering into the solid phase, we fit the low-frequency regime of the experimental DOS with the following expression:
\begin{equation}
g(\omega)=a(T)\,\omega^{b(T)}+c(T)\,.\label{fitfit}
\end{equation}
The parameter $c(T)$ relates to the self-diffusion constant in the liquid phase and vanishes in solids. 

Here, we are mostly interested in the power-law $b(T)$. In the liquid phase, as already discussed, we always find that $b(T)=1$. On the contrary, in a crystalline solid with long-range order, we have $b(T)=2$, as predicted by Debye's law. By simple extrapolation, we then expect such a power-law to interpolate between these two values by varying the temperature. 

In Fig.\ref{fig:figure3}, we show the behavior of the power-law $b(T)$ as a function of the temperature for water and Fomblin. For water, we observe a sharp jump of the power-law between the liquid value $1$ to the solid value $2$ at around $273$ K, which coincides exactly with the solidification temperature. The behavior is very drastic and it reflects the first-order nature of the liquid-solid phase transition in water. The case of Fomblin, Fig.\ref{fig:figure3}B, is more interesting. There, we observe a deviation from the liquid-like scaling below $\approx 260$ K, which is a much higher temperature than the reported pour temperature for Fomblin, $238$ K. Differently from the water-ice case, Fig.\ref{fig:figure3}A, the transition to the Debye scaling in Fomblin is not sharp, but rather continuous. This behavior is reminscent of the dynamics of the order parameter across a continuous (second-order) phase transition, \textit{e.g.}, Curie's law for magnetic materials, and it deserves further investigation. Moreover, it correlates with the different structure changes of the two systems. As shown in Fig.\ref{fig:figure7} in the Supplementary Information (SI), water (using D$_2$O, as H$_2$O does not give diffraction peaks due to the dominant incoherent neutron scattering cross sections) has a first order phase transition from a liquid state at $286$ K, represented by a broad peak in the structure factor, to a crystallized structure at $260$ K with well defined sharp peaks in the structure factor. In contrast, in the structure factor of Fomblin, there is no sharp peak but a similar broad feature appears across the whole temperature range covered. This indicates that the Fomblin structure changes gradually as a function of temperature from a liquid state (low viscosity) to a short range ordered or glassy state (high viscosity), at least in the temperature range studied here.

In order to further understand these experimental results, we turn to an analysis based on computational methods using instantaneous normal mode theory.

\section*{Instantaneous normal mode theory in a nutshell}
\begin{figure}[h]
     \centering
     \includegraphics[width=0.45\linewidth]{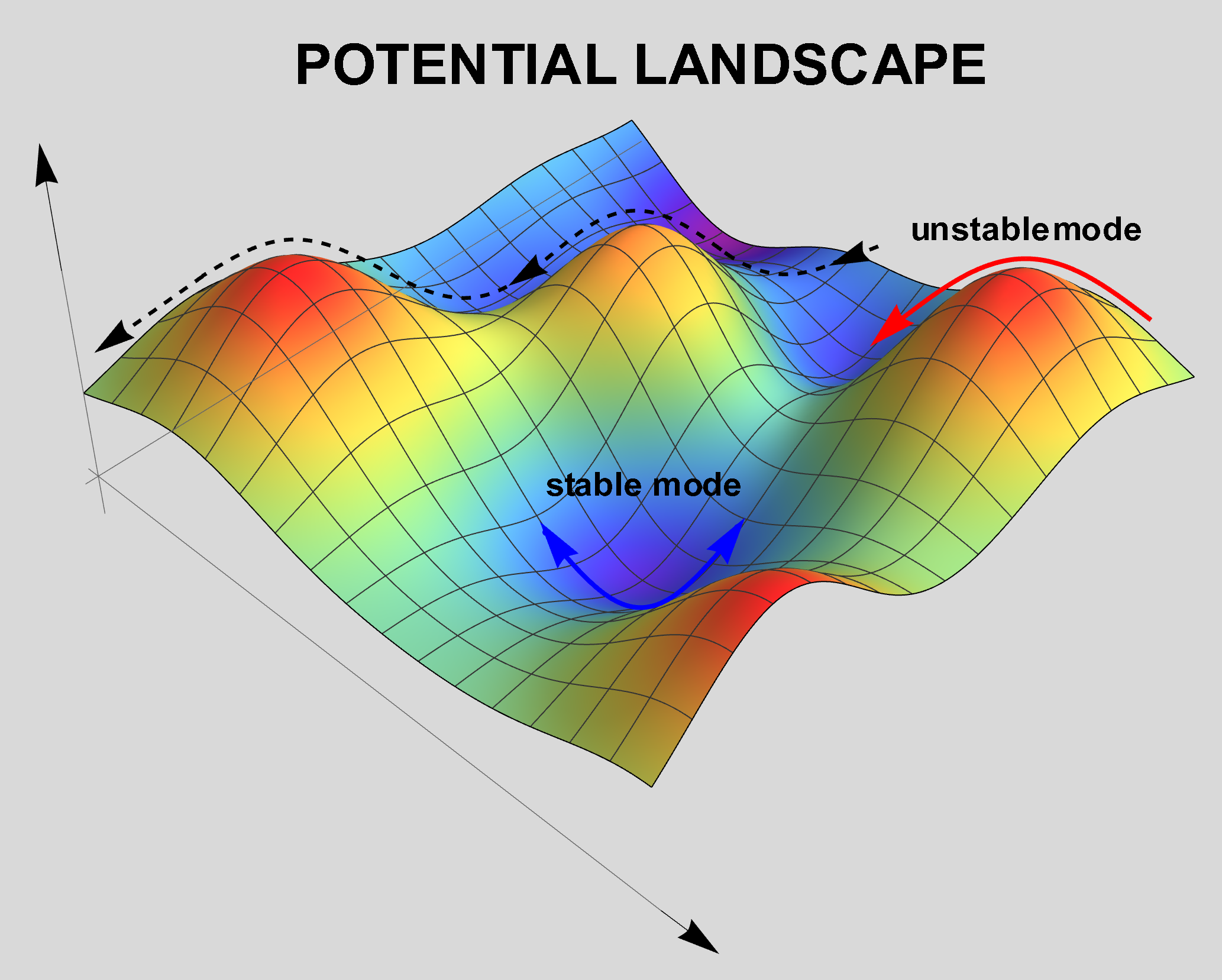}
     \caption{Schematic illustration of the typical potential energy landscape of a liquid and the corresponding stable and unstable normal modes. Regions with negative local curvature correspond to unstable modes, while minima with positive curvature correspond to solid-like stable modes performing a quasi-harmonic motion. The color-map indicates the local value of the potential energy.}
     \label{fig:0}
 \end{figure}

The dynamics of liquids are profoundly different from solids as they do not display any translational order, and are more complex than gases due to their high density and strong particle interactions. In solids, atomic motion is entirely characterized by stable oscillations around well-defined potential minima, which are well approximated by a quadratic function of the coordinates. In liquids, or more in general disordered systems, there are many minima contributing to the thermodynamics, together with negative curvature regions and saddles connecting those minima (see Fig.\ref{fig:0} for a cartoon). In the end, these hopping processes across energy barriers are the responsible for macroscopic diffusion, and cannot be neglected.

From a computational point of view, in order to derive the density of states of liquids, a common approach is to calculate the DOS using the Hessian matrix, by extending the concept of normal modes in solids to \textit{instantaneous normal modes} (INMs) (see \cite{doi:10.1021/jp963706h} for a review on the topic), and proceed with a normal-mode analysis for the liquid state \cite{doi:10.1063/1.457664,doi:10.1063/1.476768}. The main idea behind the INM approach is that, for short time-scales, a liquid is not so different from a solid, or that in more technical words, beyond the hydrodynamic regime, solids and liquids are alike (\textit{e.g.}, they both support propagating shear waves \cite{trachenko2015collective,BAGGIOLI20201}). Maxwell \cite{maxwell1867iv} suggested that the time-scale separating the liquid-like from the solid-like regime corresponds to the stress relaxation time, derived from linear viscoelasticity. As a concrete manifestation of that idea, Zwanzig \cite{PhysRev.156.190} showed that a normal mode analysis for liquids is still meaningful at short time-scales, \textit{i.e.}, at each instantaneous snapshot.

For each liquid configuration, the Hessian matrix is a $3N\times 3N$ matrix, evaluating the second derivatives of the potential energy. Its elements are constructed as follows:
\begin{align}
H_{i\mu,j\nu}(\mathbf{R})=\frac{1}{\sqrt{m_i m_j}}\frac{\partial^2V}{\partial r_{i,\mu}\partial r_{j,\nu}},
\end{align}
where $i,j=1,...,N, \mu,\nu=x,y,z$. $\mathbf{R}\equiv {\mathbf{r}_1,...,\mathbf{r}_N}$ represents each liquid configuration and $\bold r_i$ is the position of the ith atom. V is the potential energy and $r_{i,\mu}$ represents the $\mu$-coordinate of the ith atom. 
The instantaneous normal mode frequencies $\omega_i$ are the square roots of the eigenvalues of the dynamical matrix. The INM spectrum is then obtained,
\begin{align}
\left<\rho(\omega)\right>=\left<\frac{1}{3N}\sum_i^{3N}\delta(\omega _i-\omega)\right>,
\end{align}
by averaging on several instantaneous configurations, indicated by the $\langle \cdot \rangle$ symbol.

In short, INMs are the eigenvalues of the force constant matrix at an instant of time. Rhaman and collaborators \cite{mandell1976crystal} early realized that the diagonalization of such a matrix in disordered systems would give rise not only to positive eigenvalues ($\lambda>0$), but also to negative ones  which correspond to purely imaginary frequencies, $\lambda_\alpha\equiv \omega_\alpha^2<0$, and which are labelled as unstable INMs. The INM density of states can be therefore split into a stable part $g^{\text{INM}}_s(\omega)$, corresponding to the positive eigenvalues, and an unstable one $g^{\text{INM}}_u$, corresponding to the negative ones. For the unstable part, it is conventional to redefine a positive definite ``frequency'' $\tilde \omega =-i \sqrt{\lambda}$ and plot the corresponding ``density of states'' along the negative frequency axes by identifying $\tilde \omega=-\omega$. In this way, one can define a generalized INM density of states $g^{\text{INM}}(\omega)$, which corresponds for $\omega>0$ to $g^{\text{INM}}_s(\omega)$ and for $\omega<0$ to $g^{\text{INM}}_u(\omega)$. We will follow this convention. Importantly, only the stable part of the INM spectrum can be directly interpreted as a physical density of states function of real eigenfrequencies. Therefore, in the rest of this work, when we will compare the experimental DOS to the INM one, we will always refer to the stable part of the INM spectrum, $g^{\text{INM}}_s(\omega)$, or equivalently $g^{\text{INM}}(\omega)$ for $\omega>0$. In this direction, a comparison between the stable part of the INM spectrum and the experimental DOS of heavy water at $290$ K has been already shown in Ref.\cite{DAWIDOWSKI2000247}. We also notice that previous studies \cite{10.1063/1.1624057,PhysRevResearch.6.013206,Moon2024} have discussed the density of states of liquids in terms of the interplay of a gas-like and a solid-like component. In that picture, the solid-like contribution relates to the stable part of the INM spectrum.

In order to understand the distinction between positive and negative eigenvalues further, it is illustrative to think of a liquid as a collection of relatively stable local minima, around which the dynamics are harmonic and solid-like, accompanied by structural relaxation in the form of barrier crossing to neighbor wells, with a certain hopping frequency \cite{doi:10.1063/1.1672587,PhysRevA.28.2408} (see Fig.\ref{fig:0}). The imaginary frequency modes relate to these relaxational dynamics and, from a potential landscape picture, they correspond to visiting regions of the potential with locally negative curvature. In simpler words, unstable modes are a measure of fluidity, as suggested by several authors \cite{laviolette1985multidimensional,PhysRevB.33.262}. It then comes as no surprise that unstable INMs bear a close relation to diffusion \cite{doi:10.1063/1.458192,doi:10.1063/1.469947,doi:10.1063/1.468407,doi:10.1063/1.3701564,doi:10.1063/1.478211,doi:10.1063/1.475376,PhysRevE.65.026125} and several other properties of liquids \cite{doi:10.1063/1.460252,doi:10.1063/1.464106,PhysRevE.104.014103,moon2022microscopic,doi:10.1063/1.5127821,PhysRevLett.85.1464}.

The INM density of states has been investigated in several works using molecular dynamics simulations (\textit{e.g.}, water \cite{PhysRevLett.84.4605,PhysRevE.64.036102,PhysRevLett.78.2385,cho1994instantaneous}, CS$_2$\cite{doi:10.1063/1.479810}, glass-forming liquids \cite{PhysRevLett.74.936,shimada2023instantaneous}, and even proteins \cite{SCHULZ2009476}). Importantly, it has been corroborated by many simulations that at low frequency both the stable and unstable branches of the INM DOS follow a linear scaling in frequency such that the generalized INM DOS can be written as
\begin{equation}
    g^{\text{INM}}(\omega)=a(T) |\omega|+ \dots \label{inin}
\end{equation}
where the ``$\dots$'' indicate higher order corrections which are different for the stable and unstable branches and not relevant for the present discussion. To avoid clutter, we will indicate the slope of the linear regime in Eq.\eqref{inin} with the same symbol $a(T)$ used for the experimental density of states in Eq.\eqref{figdue}. Nevertheless, a priori, the two slopes are not necessarily the same, as we will explicitly confirm. We emphasize that we are interested in the generalized INM density of states in Eq.\eqref{inin} evaluated on the positive frequency axes, corresponding to the stable frequencies and positive eigenvalues.

From a theoretical perspective, several explanations for the linear scaling in Eq.\eqref{inin} have been proposed in the past, based on different degrees of simplification \cite{doi:10.1063/1.468407,doi:10.1073/pnas.2022303118,PhysRevE.55.6917,doi:10.1063/1.463375,doi:10.1063/1.465563,doi:10.1063/1.467178,doi:10.1073/pnas.2119288119,doi:10.1063/1.457564,PhysRevE.66.051110,keyes-2005}. In one way or another, all of them attribute this scaling to the presence of unstable modes. On the contrary, not much is known about the pre-factor $a(T)$ in Eq.\eqref{inin}. Simulations show that the temperature dependence of the linear coefficient $a(T)$ is not universal, and theory suggests that it does not depend on a simple physical parameter (as for the Debye's coefficient in solids), but rather on both the details of the topology of the potential landscape and the associate complex thermodynamic structure. In the literature, one can find simulated systems in which such a coefficient decreases with temperature, and systems in which it increases. In this respect, the emblematic examples are Lennard Jones liquid \cite{doi:10.1063/1.468407} for the former, and CS$_2$ for the latter \cite{doi:10.1063/1.479810}.

Within the framework of INM theory, Keyes \cite{doi:10.1063/1.468407} proposed a theoretical framework, in good agreement with the data from simulations \cite{doi:10.1063/1.468407,doi:10.1063/1.473481}, to predict the whole frequency behavior of the unstable INM density of states. In this series of works \cite{doi:10.1063/1.468407,doi:10.1063/1.479810} (see \cite{keyes-2005} for a review), Keyes and collaborators derived also a semi-analytical expression for the temperature dependence of the linear slope $a(T)$.

The theory predicts that the dominant contribution to the slope in Eq.\eqref{inin} is of the form
\begin{equation}\label{slopetheory}
    a(T)\propto e^{-\langle E\rangle/k_B T}\,,
\end{equation}
where $\langle E\rangle $ is the average barrier height at zero frequency in the potential landscape. Eq.\eqref{slopetheory} was derived for unstable modes. Nevertheless, as already explained above, the stable and unstable branches are symmetric at low frequency. Therefore, we here assume that Eq.\eqref{slopetheory} holds also for the stable part of the spectrum, that can be directly compared to the experimental results. 

The mean activation energy $\langle E\rangle$ plays a fundamental role in relating the INM properties to the temperature dependence of the self-diffusion constant, and it is ultimately connected to the hopping rate along the potential barriers. The concrete definition of $\langle E \rangle$ is subtle and not very meaningful when the distribution of the energy barriers is broad. Given the existing theories \cite{doi:10.1063/1.468407,doi:10.1063/1.479810}, the best definition of $\langle E \rangle$ that one could provide is that of an average inflection point energy, which serves as a natural zero for the barrier energy as a function of the frequency $\omega$, and which can be estimated directly in the soft potential model \cite{PhysRevE.55.6917,10.1063/5.0158089}. Because of the absence of a robust theoretical definition for the activation energy, in the rest of this manuscript we will take $\langle E \rangle$ as a phenomenological fitting parameter.

In summary, Eq.\eqref{slopetheory} is a simple formula (it neglects several microscopic details about the topology of the potential landscape) that nevertheless provides a sharp prediction. The rest of the manuscript will be devoted to validate this prediction against the experimental data. 

\section*{Instantaneous normal mode analysis}

At each temperature, we performed INM analysis for $100$ different liquid configurations generated at $5$ ps intervals during the last $500$ ps of the full simulation using GROMACS. The INM spectrum was averaged over the analyzed liquid configurations.

\begin{figure*}[!t]\centering
\includegraphics[width=0.97\linewidth]{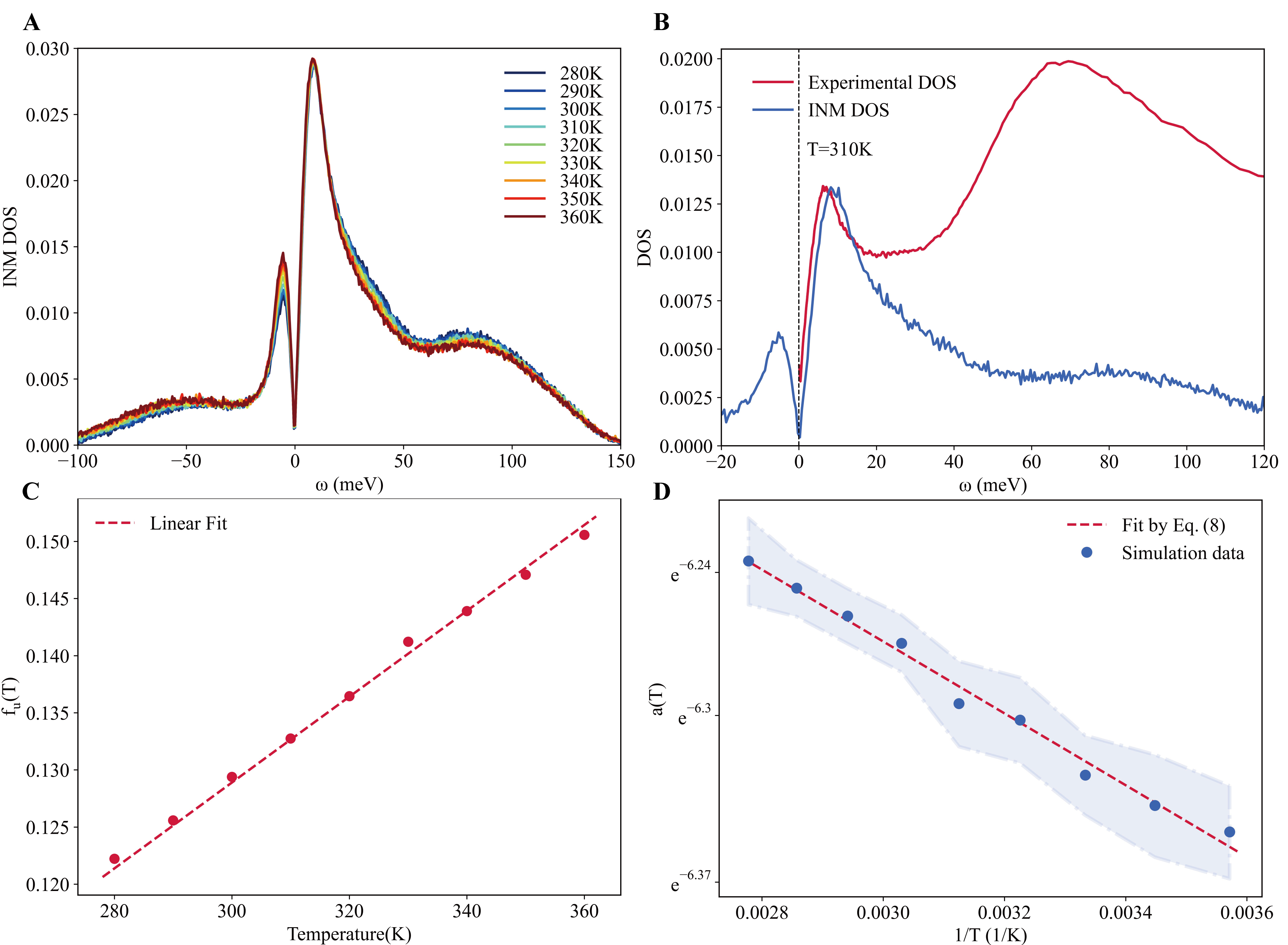} 
\caption{The instantaneous normal mode (INM) analysis for liquid water. \textbf{(A)} The INM density of states (plotted with the imaginary frequency branch on the negative $\omega$ axis) for different temperatures. \textbf{(B)} Experimental data for the DOS obtained by INS (red), and INM density of states (blue) at $310$ K. The data are normalized using the height of the first peak. \textbf{(C)} The fraction of unstable modes $f_u(T)$ as a function of temperature. The dashed line indicates the result of a linear fit. \textbf{(D)} The slope of the INM DOS $a(T)$ as a function of the temperature in a linear logarithmic scale. The line indicates the fit to the prediction from INM theory, Eq.\eqref{fitfit}. The background colored regions illustrate the uncertainties of the numerical data.}
\label{fig:figure6}
\end{figure*}

Unfortunately, for Fomblin we have not been able to carry out an INM analysis since the force potential is unknown. For liquid water, we carried out a $1$ ns molecular dynamics simulation of the flexible TIP3P model. Previous normal mode analyses for supercooled water can be found in \cite{cho1994instantaneous,PhysRevE.64.036102,PhysRevLett.78.2385,PhysRevLett.84.4605}. The resulting INM density of states after average, including both the stable branch $g^{\text{INM}}_s(\omega)$ and the unstable one $g^{\text{INM}}_u(\omega)$, is shown in Fig.\ref{fig:figure6}A. We have followed the standard notation and plotted the imaginary frequencies, corresponding to the unstable modes, on the negative frequency axis. As evident from Fig.\ref{fig:figure6}A, both the stable and unstable parts display a clear linear-in-frequency behavior at low frequency. Moreover, the slope is the same for the two parts \cite{keyes-2005}, as anticipated. In other words, both the stable and unstable branches of the INM DOS follow the linear behavior presented in Eq.\eqref{inin}.

Importantly, only the stable branch of the INM DOS, $g^{\text{INM}}_s(\omega)$, can be compared to the experimentally measured DOS $g(\omega)$. The density of states for stable INMs, $g^{\text{INM}}_s(\omega)$ can be decomposed into two regimes. The region below $50$ meV involves mostly modes related to the translational motion. On the contrary, the region above $50$ meV is governed by rotational modes \cite{cho1994instantaneous}. By increasing temperature, the number of unstable modes increases. Additionally, despite both parts of the INM DOS displaying a clear peak at low frequency, the one for stable modes appears to be insensitive to the temperature $T$, while the one for the unstable modes grows in intensity with temperature. This is a concrete proof that the two branches are symmetric only in the low-frequency linear regime presented in Eq.\eqref{inin} but not for larger frequencies.

For completeness, in SI, we show the comparison between our results for the INM DOS of water at $300$K and the existing data in the literature that used TIP4P/2005 rigid water potential \cite{KUO2023140612} and flexible SPC water potential \cite{huang2013localization}. Evidently, the low-frequency behavior of the INM DOS, and in particular the linear scaling and its slope, do not depend on the specific potential used. Nonetheless, the higher energy dynamics do show strong dependency on different potentials. This will be a topic for further studies.

In Fig.\ref{fig:figure6}B, we show the comparison of the experimental DOS for water at $310$ K and the DOS of stable INMs at the same temperature. The two curves have been normalized to the first peak. As evident from Fig.\ref{fig:figure6}B, there is a crucial difference between the two curves since the INM density of states $g^{\text{INM}}_s(\omega)$ does not contain the diffusion component, and therefore $g^{\text{INM}}_s(0)=0$, while the experimental curve clearly shows a finite value at zero frequency. As explained above, the zero frequency value is given by the self-diffusion coefficient $D$ and increases with temperature. In SI, we show a more in depth comparison between the two curves at different temperatures. Since $g(0)$ diminishes when the temperature is lowered, the two curves resemble each other more near the solid-phase. Indeed, in a low-temperature solid, we expect the two curves to be identical (see \cite{moon2022microscopic} for a similar observation). Moving towards higher frequency, we observe that in the low-energy regime, below approximately $20$ meV, the two curves are similar to each other. Both of them show a clear linear-in-frequency regime, and then a sharp peak with a similar linewidth. At larger frequencies, the two curves differ considerably. It is plausible that this mismatch is caused by the relative intensity of the features in the neutron scattering DOS which is skewed by the larger scattering cross section of hydrogen atoms compared to oxygen (see Methods for details). Another possible reason may be that the INM approach itself, or the potential chosen, can not reproduce the high energy modes properly. 

%\begin{SCfigure*}[\sidecaptionrelwidth][t!]

A crucial quantity in the INM analysis is given by the fraction of unstable modes $f_u$, which is defined as the ratio of the number of unstable modes to the total 3N modes. Fig.\ref{fig:figure6}C displays the fraction of unstable modes $f_u$ as a function of temperature. As expected, the fraction increases with increasing temperature and for the case of water exhibits an evident linear in $T$ dependence. Interestingly, this functional form coincides exactly with the prediction of a specific theoretical model discussed in the literature, the ``random energy model'' \cite{PhysRevE.66.051110} (\textit{e.g.}, Eq.13.33 in \cite{keyes-2005}).

Let us now focus on the linear slope $a(T)$, which is plotted in Fig.\ref{fig:figure6}D as a function of the inverse temperature. First, the behavior of the linear slope is consistent with the result from the experiments, as it increases monotonically with the temperature. In order to provide a more quantitative understanding of the slope, we resort to Eq.\eqref{slopetheory} based on the INM analysis. As shown in Fig.\ref{fig:figure6}D, the theoretical formula is in good qualitative agreement with data from simulations. In particular, by plotting the simulation data in a linear logarithmic scale in Fig.\ref{fig:figure6}D, we unequivocally show that the slope follows an Arrhenius-like exponential form $\exp\left(-\langle E\rangle/k_B T\right)$, as predicted by INM theory, \textit{i.e.}, Eq.\eqref{slopetheory}. Given the complexity of liquid dynamics, it is remarkable that INM theory is able to capture the qualitative temperature dependence of $a(T)$. In order to provide a more quantitative analysis, we fit our data for the slope $a(T)$ using Eq.\eqref{slopetheory} with an additional constant pre-factor $A$,
\begin{align}
    &a(T)=A \,e^{-\langle E\rangle/k_B T}.\label{fitfit}
\end{align}
According to our best fit, we obtain the following values for the various parameters
\begin{align}
    & A= \left(2.98\times 10^{-3} \pm 5.48\times 10^{-5}\right)\,\text{ps}^2,\nonumber\\
    &\langle E\rangle= \left(12.99 \pm 0.51 \right)\,\text{meV}.\label{fitinm}
\end{align}
Next, we will extend the analysis to the experimental data and compare the outcomes with the results from the simulations performed in this section.

\section*{Comparing the experimental DOS to the stable INM density of states}
\begin{figure}[h]\centering
\includegraphics[width=0.97\linewidth]{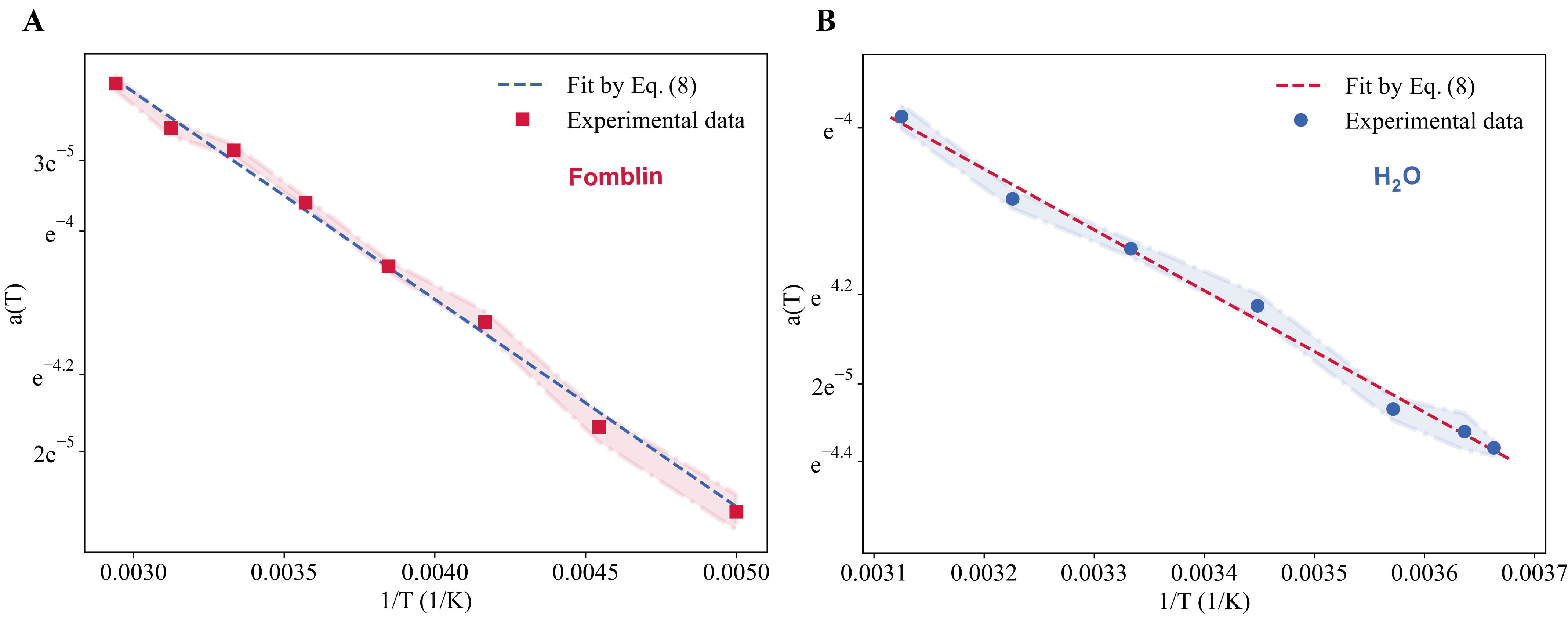}
\caption{The temperature dependence of the linear slope $a(T)$ for the experimental DOS of water \textbf{(A)} and Fomblin \textbf{(B)} in a linear-log plot. The background shaded regions indicate the uncertainties of the experimental data. The colored lines indicate the fit with Eq.\eqref{fitfit}.}
\label{fig:figure2}
\end{figure}

We return to the experimental data for the DOS of water and Fomblin displayed in Fig.\ref{fig:figure1}. Here, we are mostly concerned with the temperature dependence of the slope for the linear scaling which has been observed to be universal in the whole liquid phase. The data, obtained by fitting the experimental results with Eq.\eqref{fitfit}, are shown in Fig.\ref{fig:figure2}. For Fomblin, the slope is extracted by fitting the experimental data from INS normalized by the total area. This normalization is justified by the fact that around $\approx 120$ meV, where our numerical data stops, the spectral weight is already very small (see Fig.\ref{fig:figure1}B), and the remaining tail is negligible. For water (see Fig.\ref{fig:figure1}A), this is not the case, as around $\approx 140$ meV, the DOS is still large. As a consequence, a normalization of the DOS by the area of the curves up to that experimental cutoff would lead to uncontrollable results. Therefore, for water, we have normalized all the curves by their value at zero frequency, using the experimental and simulation data for the self-diffusion constant $D(T)$ \cite{doi:10.1080/08927022.2018.1511903}, which is known precisely at all temperatures. This problem with the normalization of the experimental data for water will not affect qualitatively our findings and our analysis, but it will make impossible to reliably study quantitative features such as the value of $\langle E \rangle$. On top of that, the experimental DOS for water is also affected by the different scattering cross sections of different atoms. This implies that certain vibrational modes can be over- or under-represented if there is a large difference in scattering power between the elements involved in different modes as it happens for hydrogen and oxygen (see Methods).

Let us first focus on the qualitative functional dependence of $a(T)$ that can be analyzed using the data presented in Fig.\ref{fig:figure2}.
For both systems, the slope of the experimental DOS increases monotonically with temperature. Additionally, we find that both liquids display an exponential Arrhenius-like behavior as in Eq.\eqref{slopetheory}, which is derived from INM theory \cite{doi:10.1063/1.468407,doi:10.1063/1.479810} based on the dynamics of unstable modes. 

\begin{figure}
    \centering
    \includegraphics[width=0.6\linewidth]{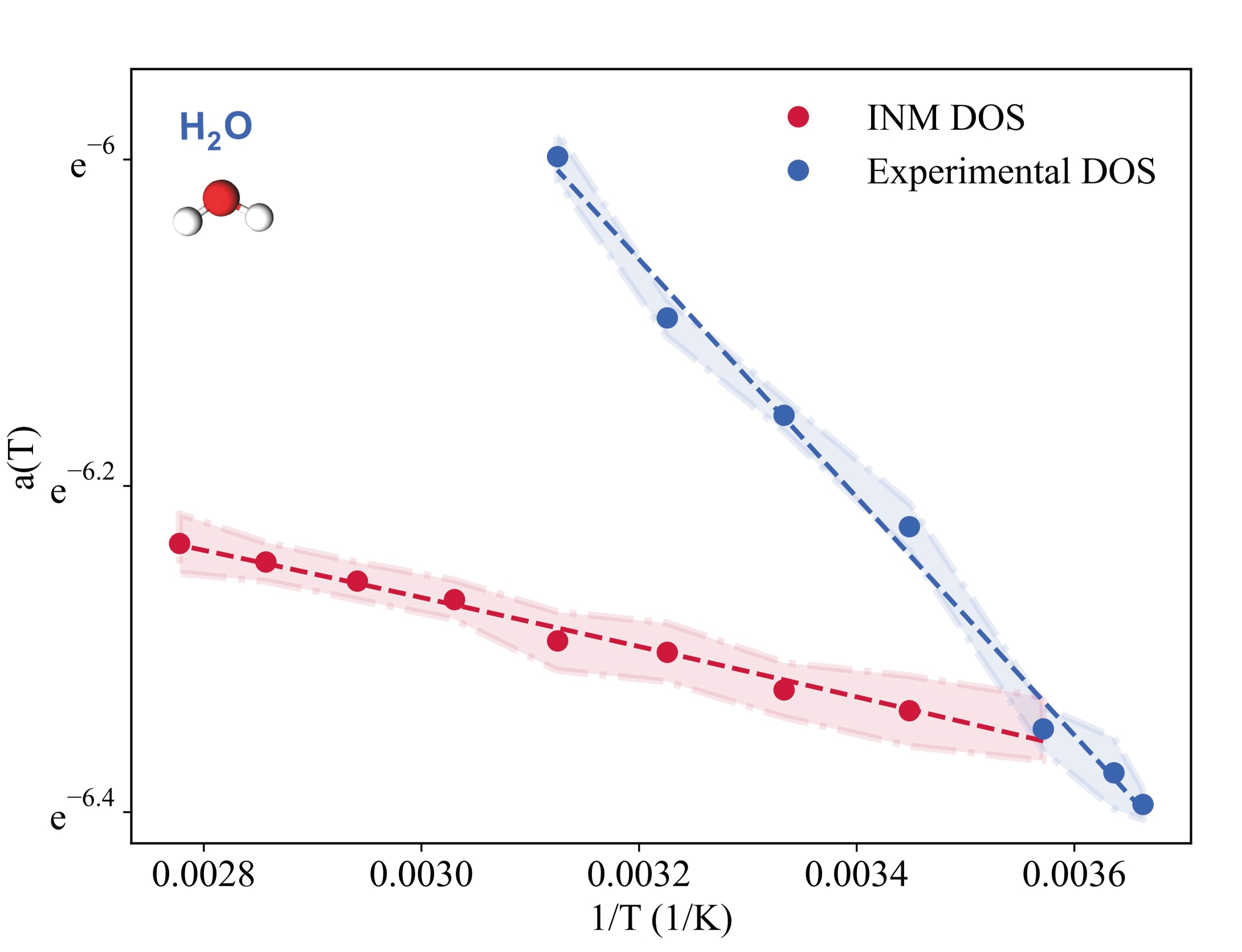}
    \caption{A direct comparison of the slope of the low-frequency DOS $a(T)$ extracted from the experimental data (blue color) and the INM analysis (red color). The dashed lines emphasize the common exponential behavior $\exp\left(-\langle E\rangle/k_B T\right)$. The color shaded regions indicate the error bars in the data.}
    \label{fig:last}
\end{figure}

Taking into account the difficulties previously discussed, we attempt a more quantitative analysis by fitting the slope of the experimental DOS of water using Eq.\eqref{fitfit}. We obtain the following results:
\begin{align}
    & A= (0.18\pm 0.02)\,\text{ps}^2,\\
    &\langle E\rangle^{\text{exp}}= \left(62.83\pm 2.36\right) \,\text{meV}.
\end{align}
By comparing these results with those obtained from the INM DOS, and presented in Eq.\eqref{fitinm}, the discrepancy between the two set of values is evident. First, we notice that the fitted parameters concerning the experimental data are inevitably affected by much larger errors. Then, the overall prefactor $A$ is almost two orders of magnitude different. This is not a surprise as the latter is highly sensitive to the normalization of the DOS, and therefore its value not reliable for the reasons explained above. Additionally, we notice that the energy scale $\langle E\rangle$ extracted from the experimental data is $\approx 5$ times larger than that from the simulation data. This is made more evident by presenting together the results for the slope $a(T)$ from the experimental data and the INM theoretical analysis in Fig.\ref{fig:last}. In summary, both the experimental and simulation data show a clear exponential behavior, $\exp\left(-\langle E\rangle/k_B T\right)$, but with a different activation energy. We believe that this discrepancy is due to the strong anharmonic effects that are inevitably present in liquids and that are not entirely captured by analysis of the Hessian eigenvalues on which INM theory is based.

In order to conclude our quantitative analysis, we analyzed the experimental data for the slope $a(T)$ for Fomblin following the same method. We obtain the following values for the fitting parameters:
\begin{align}
    &A= (5.29\pm 1.63)\,10^{-3}\,\text{ps}^{2} \\
    &\langle E\rangle^{\text{exp}}=\left(24.91 \pm 0.73 \right) \,\text{meV}.
\end{align}
These results suggest that the energy scale $\langle E\rangle$ for fomblin is about a factor $2$-$3$ smaller than for water. At this point, without performing a more comprehensive analysis involving more liquid systems, it is hard to make any claim about this energy scale and its physical meaning. We plan to extend our investigation in the near future by performing large-scale MD simulations and INM analysis of several liquid substances.

Without further analysis, we can conclude that our experimental results confirm the validity of the universal linear law for the DOS of liquids $g(\omega)\propto \omega$ in a wide range of temperature in the liquid phase. On top of that, and as the main novelty of our work, our results reveal that the temperature dependence of the slope associated to this universal scaling law exhibits an exponential $\exp(-\langle E\rangle/T)$ form. Importantly, this exponential behavior can be derived using INM theory that successfully reproduces the qualitative trend of the experimental data.

However, we find that the INM analysis is not sufficient from a quantitative point of view. In particular, as evident from Fig.\ref{fig:last}, it cannot correctly capture the precise value of the energy scale $\langle E\rangle$, whose precise meaning remains unclear. We leave this important point as an open question for the future.

\section*{Discussion}
In this work, we studied the DOS of water and Fomblin oil, combining experimental neutron scattering techniques (INS), MD simulations and INM theory. Our focus is the low-energy regime of the DOS and in particular (I) its scaling with frequency, and (II) its temperature dependence. In such a regime, and independently of the value of the temperature (from the melting temperature to $\approx340$ K), we have experimentally verified that both liquids display a universal linear in frequency scaling $g(\omega)= a(T) \omega+\dots$, which was previously experimentally observed for only one value of temperature in Ref.\cite{doi:10.1021/acs.jpclett.2c00297}, and predicted before in Ref.\cite{doi:10.1073/pnas.2022303118}.

By analyzing the temperature dependence of the linear slope $a(T)$, we observed that $a(T)$ grows monotonically with temperature over a wide range of temperatures in the liquid phase, and that the functional behavior is compatible with an exponential Arrhenius-like form typical of thermally activated dynamics. In order to rationalize this behavior, we have resorted to INM theory \cite{doi:10.1063/1.468407}. Our INM simulations show a good qualitative agreement with the experimental data and predict the same temperature dependence of the slope values as observed in the INS experiments. In particular, the experimental data confirmed that the temperature dependence of the linear slope is dominated by an exponential factor $\exp(-\langle E \rangle /k_B T)$, as predicted by INM theory \cite{doi:10.1063/1.468407,doi:10.1063/1.479810,keyes-2005}. Nevertheless, the INM analysis fails in reproducing quantitatively the experimental results (see Fig.\ref{fig:last}), and more work has to be done to establish the role and the physical meaning of the energy scale appearing in the exponential behavior of the slope as a function of temperature. We propose that this discrepancy originates from the anharmonic effects that are neglected in the INM analysis, based on the instantaneous Hessian matrix.

Finally, we have tracked the power-law scaling of the low-frequency portion of the experimental DOS across the melting temperature. For water, we have consistently observed a sharp transition between the liquid-like linear scaling to the quadratic Debye law, which appears around the expected melting temperature. On the contrary, for Fomblin, the transition from the liquid behavior to Debye's law is continuous, as expected from its polymeric glassy structure. These findings imply that the low-frequency temperature dependent behavior of the experimental DOS is able not only to capture the solid-liquid transition but also its nature, whether a sharp first-order thermodynamic transition or a continuous glassy-like transition.

In conclusion, we hope that our analysis will motivate future studies on the low-frequency behavior of the liquid DOS, a fundamental but rather unexplored topic that deserves further attention.

\section*{Methods}\label{app1methods}
\subsection*{Samples}
Two samples have been studied with INS. General-purpose laboratory grade of deionized water with a mass of $0.5$ g is filled inside an annular aluminum can having $0.1$ mm gap. A similar can with $0.5$ mm gap is filled with $4$ g of Fomblin oil (YL $25/6$, Fomblin\textsuperscript{\textregistered} Y LVAC $25/6$ Solvay) \cite{test}; CF$_3$O[-CF(CF$_3$)CF$_2$O-]$_x$(-CF$_2$O-)$_y$CF$_3$ where $x/y$ gives an average molecular weight of $3,300$. These aluminum cans have been specially designed with high strength aluminum to maintain the integrity of the sample can under high vapor pressure of water at high temperatures. A half dozen of sample cans have been destroyed during the test. The estimated transmission is $90\%$ for these samples and this ensures the $10\%$ level of scattering to minimize multiple scatterings.
\subsection*{Inelastic neutron scattering.}
The inelastic neutron scattering measurements were conducted using Pelican – the time-of-flight cold neutron spectrometer at the Australian Nuclear Science and Technology Organisation. A neutron energy of $3.7$ meV (wavelength of $4.69$ Å) was employed, yielding an energy resolution of $0.135$ meV at the elastic line. Sample temperature control, ranging from $1.5$ K to $800$ K, was achieved via a top-load cryo-furnace. Background subtraction involved measuring the corresponding empty can under identical conditions as the samples. Additionally, a standard vanadium sample was measured for detector efficiency normalization and energy resolution determination. The experimental dynamic scattering function $S(Q,\omega)$ was derived using the Large Array Manipulation Program (LAMP) \cite{lamp}, following background subtraction and detector normalization. For density of states (DOS) determination, the $S(Q,\omega)$ corresponding to the neutron energy gain side was utilized.
The DOS for the solid phase of materials was determined through a standard procedure outlined in various textbooks. In the incoherent one-phonon approximation, the relationship between DOS and the experimental scattering function for a Bravais powder sample (isotropic system) can be expressed as \cite{boothroyd2020principles}:
\begin{align}
g(\omega)\, = \,C \frac{\omega}{{Q}^2}\,S(Q,\omega)\,\left(1-e^{-\hslash\omega/k_B T}\right).\label{ins1}
\end{align}
Here, $C$ is a factor containing the atomic mass and Debye-Waller factor, $\exp\left(-2W\right)$, which is taken as unity for all samples. Furthermore, $k_B$ is the Boltzmann constant. Experimentally, the DOS is averaged over the $Q$ range covered, typically from $0.2 \text{Å}^{-1}$ to $9 \text{Å}^{-1}$, depending on the energy range considered.

For non-Bravais samples, the measured phonon density of states (DOS) is determined by the neutron-weighted phonon density of states, known as the generalized phonon density of states (GDOS) \cite{boothroyd2020principles}:
\begin{equation}
    g_{NW}(\omega)=\sum_i f_i \frac{\sigma_i}{M_i}g_i(\omega),
\end{equation}
where the sum over $i$ includes all elements in the sample, $f_i$ is the $i_{th}$ atomic concentration, $\sigma_i$ is the total neutron bound cross section accounting for both coherent and incoherent scattering processes, $M_i$is the atomic mass, and $g_i\left(\omega\right)$ is the partial DOS of the element $i$. In the case of water, the measured DOS is predominantly influenced by hydrogen due to the higher $\sigma/M$ ratio for H compared to O. 

The experimental determination of DOS is subject to weighted averaging effects and potential errors introduced by multiple scatterings and multi-phonon scattering. To mitigate multiple scattering effects, sample thicknesses were chosen to yield approximately $10\%$ neutron scattering. For instance, sample thicknesses of $0.2$ mm and $1.0$ mm were utilized for water and Fomblin, respectively. Multi-phonon scattering effects are generally small, particularly within the low Q and low energy ranges interested in this study. The Q range covered for the low energy (< $3$ meV) DOS is from $0.2$ $\text{Å}^{-1}$ to about 2 $\text{Å}^{-1}$. Consequently, the reported GDOS here has no correction for multiple scatterings and multi-phonon scattering effects. Furthermore, the correct Debye $\omega^2$ law obtained for solid phase of the materials has indicated that these effects are at the negligible level. 

For the liquid phase, a similar procedure was followed to derive the DOS. In the low-energy limit, Eq.\ref{ins1} simplifies to:
\begin{align}
g(\omega)\, = \,C \frac{\omega^2}{{Q}^2}\,S(Q,\omega)\,\label{ins2}
\end{align}
It is noteworthy that Eq.\ref{ins2} is essentially equivalent to the formula for determining the frequency spectrum \cite{citekeyincollection} or velocity frequency function
\cite{1978Introduction} for liquids in the limit of $Q \rightarrow 0$. This same formulation was also utilized to derive the proton density of states of water \cite{chen1984hydrogen}.

%It is pertinent to consider the potential contribution from diffusion processes in liquid systems. Under the approximation of Eq.\ref{ins2}, at low energies for a given $Q$, we have:
%\begin{align}
%g(\omega)\, \propto \, \omega^2\,S(Q,\omega)\
%\end{align}

%Since $g(\omega)$ exhibits a linear relationship with $\omega$ as observed experimentally,$S(Q,\omega)$ should behave as $1⁄\omega$. For the diffusion process, the $S(Q,\omega)$ is normally modeled by a Lorentzian function, which behaves as $1⁄\omega^2$ away from the centre of the function. This would imply a constant contribution to the DOS decreasing much faster than $1⁄\omega$ with increasing energy. Hence, it is reasonable to assume that the relaxation process due to diffusion insignificantly affect the linear scaling of the DOS for liquids above certain energies.  

\subsection*{Molecular dynamics (MD) simulations.}
The MD simulations to simulate liquid water at different temperatures were performed with GROMACS \cite{van2005gromacs}. We used the flexible TIP3P water model defined in GROMACS, which allows the stretching of the O-H bond and the bending of the H-O-H angle. The flexible water was applied into the initial topology and the SETTLE algorithm was not applied. We simulated $392$ water molecules in our liquid system which was equilibrated in NVT and NPT ensembles, using the Nos\'e-Hoover thermostat and Parrinello-Rahman barostat to control the temperature and pressure. We carried a $1$ ns production MD simulation for data collection at atmospheric pressure and constant temperature. During the whole simulation, the periodic boundary conditions were employed. Using the Leapfrog-Verlet algorithm, the time step was set to $0.1$ fs. The long-range electric interactions were calculated by using the particle mesh Ewald (PME) method. The self-diffusion coefficient of the system can be determined through the mean squared displacement (MSD) via the Einstein relation:
\begin{align}
D = \frac{1}{6}\frac{d}{dt}\left<\left|\mathbf{r}_i(t)-\mathbf{r}_i(0)\right|^2\right>,
\end{align}
where $\bold{r}_i(t)$ is the position vector of the ith particle at time t.
\bibliography{sample}

\section*{Acknowledgements}

We would like to thank H. Xu, J. Douglas, Y. Feng and especially T. Keyes for fruitful discussions and related collaborations on the topic of liquids. We are grateful to Tom Keyes for comments and suggestions on a preliminary version of this manuscript. M.B. acknowledges the support of the Shanghai Municipal Science and Technology Major Project (Grant No.2019SHZDZX01) and the sponsorship from the Yangyang Development Fund. D. Y., C. S. and R. M. acknowledge the beam time awarded from ANSTO for the access to Pelican instrument (P13964).

\section*{Author contributions statement}

D.Y., C.S. and R.M. performed the experimental measurements; M.B., D.Y conceived the idea of this work; S.J, X.F. implemented the MD simulations and the INM analysis; S.J., X.F., Y.Y. and C.S. performed the analysis of the experimental and simulation data; M.B. and S.J. wrote the manuscript with the help of D.Y.

\clearpage
\newpage
\section*{Supplementary information}
\subsection*{Fitting the experimental data}
\label{app2}
\begin{figure*}[!t]\centering
\includegraphics[width=0.8\linewidth]{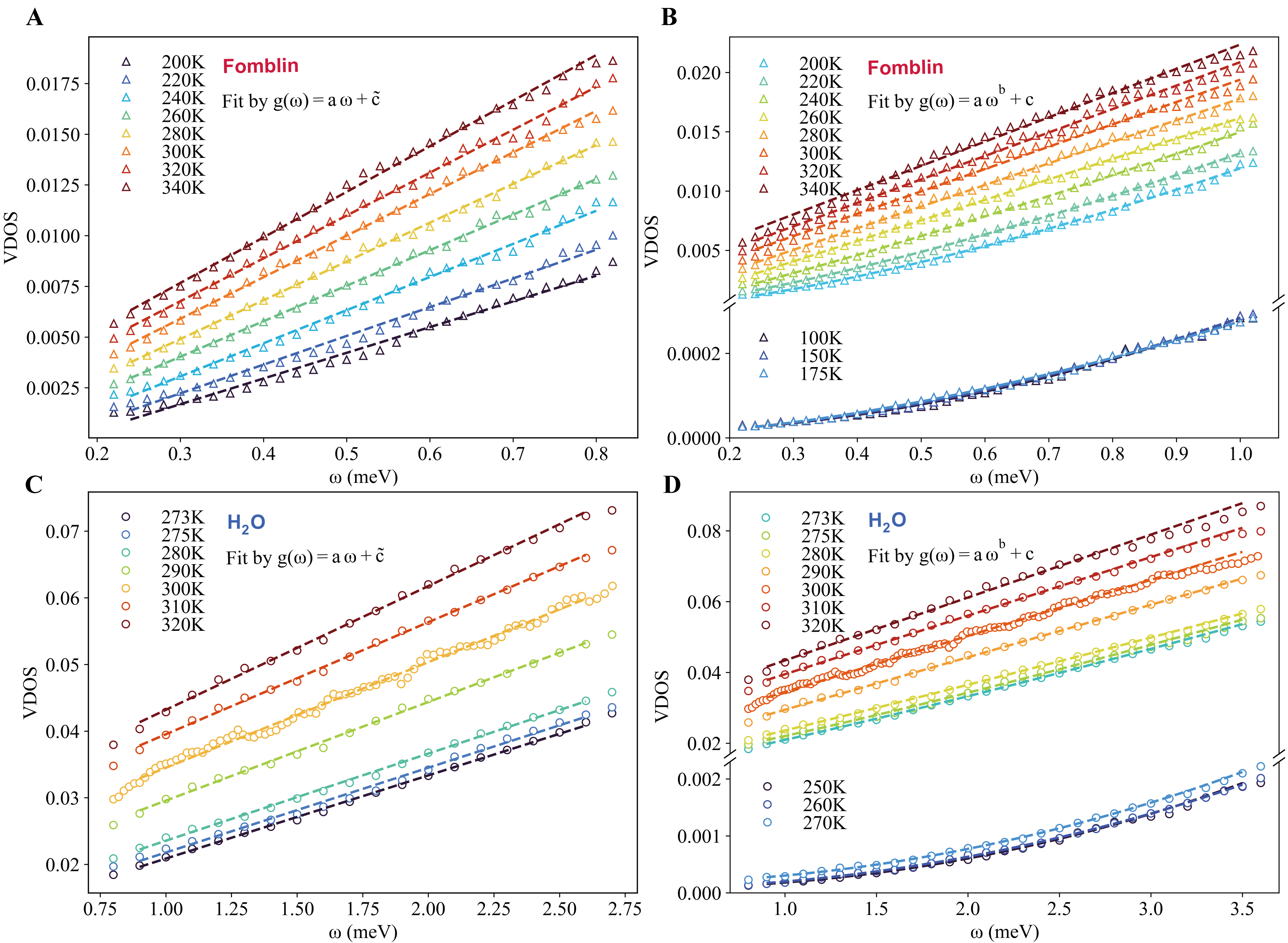}
\caption{Fits of the experimental data. The dashed lines indicate the result of the fits. \textbf{(A)} The linear fit of the area normalized experimental DOS of Fomblin oil. \textbf{(B)} The power law fit of the experimental DOS of Fomblin oil. For $T \geq 200$ K, the DOS is normalized by the area. For $T<200$ K, the data is not normalized because the experimental data stops around $20$meV. \textbf{(C)} The linear fit of the normalized experimental DOS of liquid water ($T \geq 273$ K). \textbf{(D)} The power law fit of the experimental DOS of liquid water. For $T \geq 273$ K, the DOS is normalized by the value at zero frequency. For $T<273$ K, the DOS is normalized by the area.}
\label{fig:figure5}
\end{figure*}

The linear fit of the experimental DOS of Fomblin oil, used to extract the slope presented in the main text, is plotted in Fig.\ref{fig:figure5}A. The DOS curve is normalized by the total area under the curve. On the other hand, Fig.\ref{fig:figure5}B presents the power law fit of the low frequency DOS of Fomblin. Since the value of the power extracted from the fit is not affected by the normalization of the data, the DOS data for $T<200$ K are also shown but they are not normalized to the total area because the experimental data stops around $20$ meV. 

For the linear fit of the low-frequency experimental DOS of liquid water ($T\geq 273$ K), the normalization by the area of the curves could lead to uncontrollable results due to the large spectral weight above the experimental data cutoff, $140$ meV (see Fig.\ref{fig:figure1}A in the main text). In order to normalize the DOS properly, we resorted to a different method and used the value of the DOS at zero frequency, which, in the Brownian motion approximation, is theoretically given by the self-diffusion constant $D$,
\begin{equation}\label{eee}
g(0)=\frac{2m\,D}{k_BT}.
\end{equation}
Since the exact zero frequency value of the DOS is not experimentally accessible, first, we fitted the low-frequency DOS of liquid water using
\begin{equation}\label{linear}
    g(\omega)= a(T)\,\omega+\tilde c(T),
\end{equation}
and extracted the constant term $\tilde c(T)$. By multiplying the experimental DOS with the ratio $c(T)/\tilde c(T)$, where $c(T)$ is given in terms of the self-diffusion constant $D$ as in Eq.\eqref{eee}, we have normalized all the curves by their value at zero frequency. The self-diffusion coefficient is obtained from the MD simulation, as presented in the Methods. The linear fit of the low-frequency normalized DOS of liquid water is plotted in Fig.\ref{fig:figure5}C. Finally, Fig.\ref{fig:figure5}D presents the power law fit of the low frequency DOS of water. For $T<273$ K, the DOS is normalized by the area as the self-diffusion constant vanishes in solids. We emphasize that, due to the limited range of the experimental data, the results extracted from the fits might be affected by inaccuracies and the precise quantitative numbers have to be considered with caution.

\subsection*{Verification of the crystalline, liquid and glassy phases}
\label{app3}
\begin{figure}[!t]\centering
\includegraphics[width=0.8\linewidth]{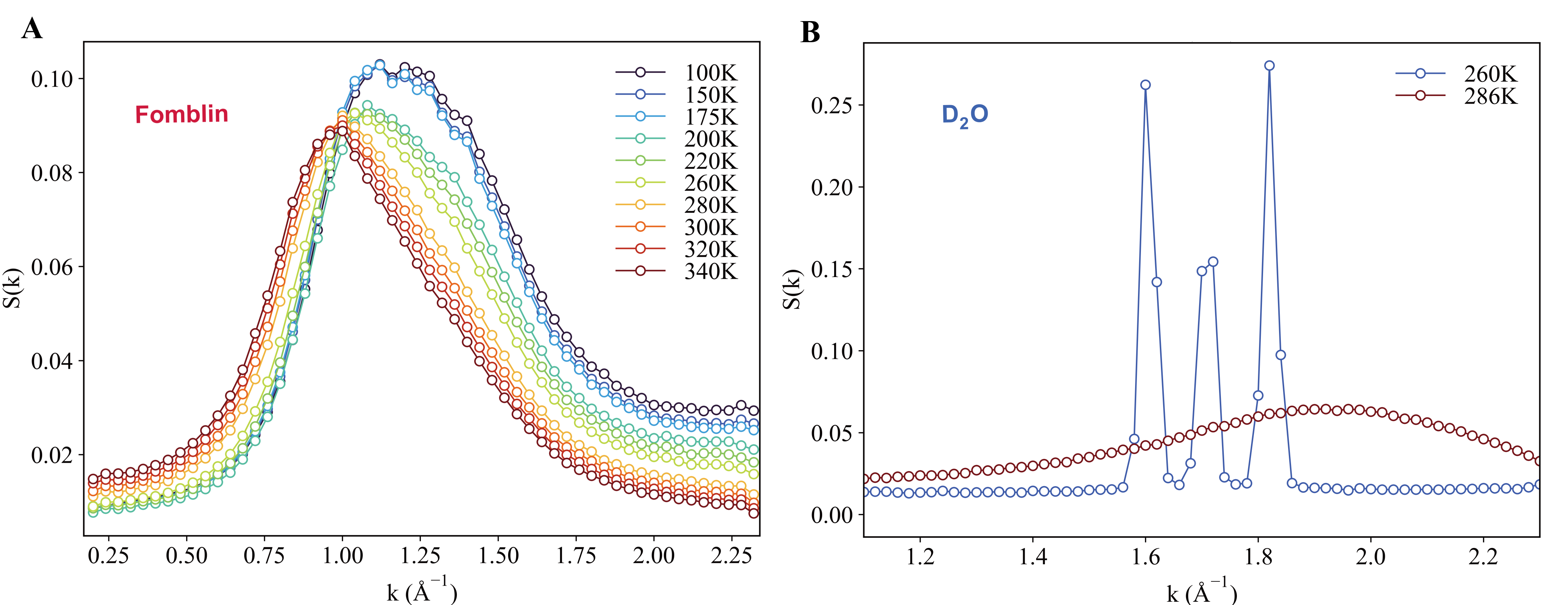}
\caption{The experimental structure factor $S(k)$ of Fomblin oil \textbf{(A)} and D$_2$O \textbf{(B)}.}
\label{fig:figure7}
\end{figure}

In Fig.\ref{fig:figure7}, we show the measured structure factor of Fomblin and water respectively. For Fomblin, the change of structure factors in a wide temperature range is shown in Fig.\ref{fig:figure7}A. For water, the structure factor of D$_2$O is presented, as H$_2$O does not give clear diffraction peaks due to the dominating incoherent neutron scattering cross sections. Fig.\ref{fig:figure7}B shows a clear first order phase transition from a crystallized structure at $260$ K, with well defined sharp peaks in the structure factor, to a liquid state at $286$ K, with a broad peak in the structure factor. In contrast, the structure factor of Fomblin does not exhibit any sharp peak for all the temperatures considered, but displays a similar broad peak shifting toward lower $k$ with increasing temperature. The continuous behavior of the structure factor of Fomblin, upon moving to the low-temperature liquid state, verifies that Fomblin is not a is not a crystalline solid but it rather exhibits a short-range glassy structure with high viscosity at low temperature. This is consistent with the drastically different behavior of the low-frequency power-law of the DOS reported for the two systems in Fig.\ref{fig:figure3} in the main text.

\subsection*{Comparison of INM data with previous literature}
\label{app6}
\begin{figure}[!t]\centering
\includegraphics[width=0.6\linewidth]{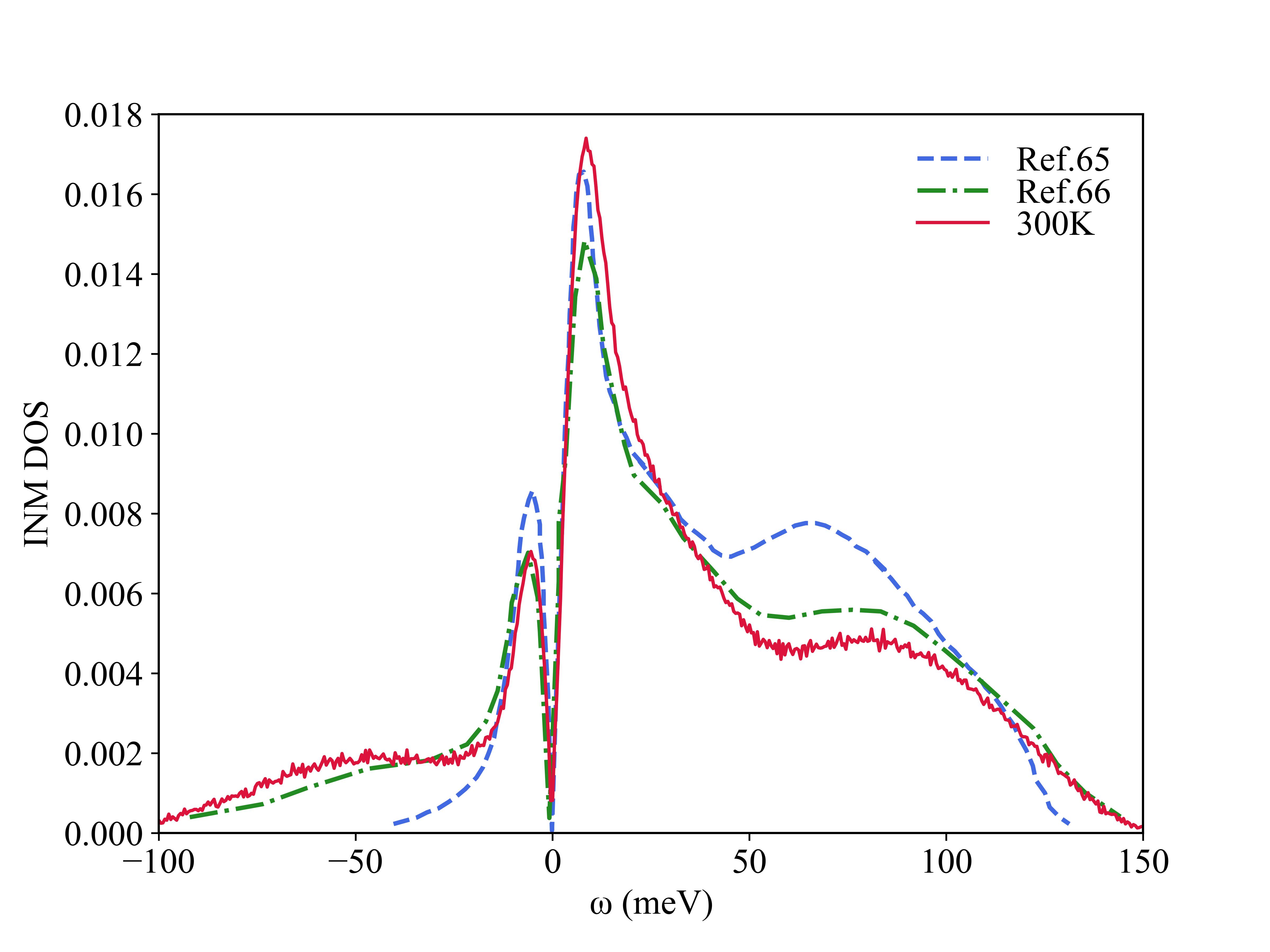}
\caption{The comparison of our INM DOS with the existing data in the literature. The blue and dashed curve is the INM DOS of TIP4P/2005 rigid liquid water at 300K from Fig.3 in Ref. \cite{KUO2023140612}. The green dash-dotted curve is the INM DOS of flexible SPC liquid water at 298K from Fig.1 in Ref. \cite{huang2013localization}. The red solid curve corresponds to the data at 300K in Fig.\ref{fig:figure6}A. All the data are normalized by the area.}
\label{fig:figure10}
\end{figure}

For completeness, we compare our results for the INM DOS of water with the other existing data in the literature. In Fig.\ref{fig:figure10}, we directly compare our data (Fig.\ref{fig:figure6}A) with those obtained using  TIP4P/2005 rigid water potential at $300$K from Fig.3 in Ref.\cite{KUO2023140612} and those obtained using a flexible SPC water potential at $298$K from Fig.1 in Ref.\cite{huang2013localization}.

This comparison reveals that the low-frequency regime is independent of the potential used and the same linear scaling $g(\omega)=a(T) \omega$ is observed. On the contrary, the DOS at higher energy depends on the potential used. Our data are very close to those obtained for flexible SPC water potential but they present marked differences with the data obtained using TIP4P/2005 rigid water potential. In particular, the DOS obtained with the rigid potential displays a much smaller number of unstable modes and a sharper peak around $65$ meV.

\subsection*{Experimental DOS versus INM DOS}\label{app5}
\begin{figure*}[h]\centering
\includegraphics[width=0.97\linewidth]{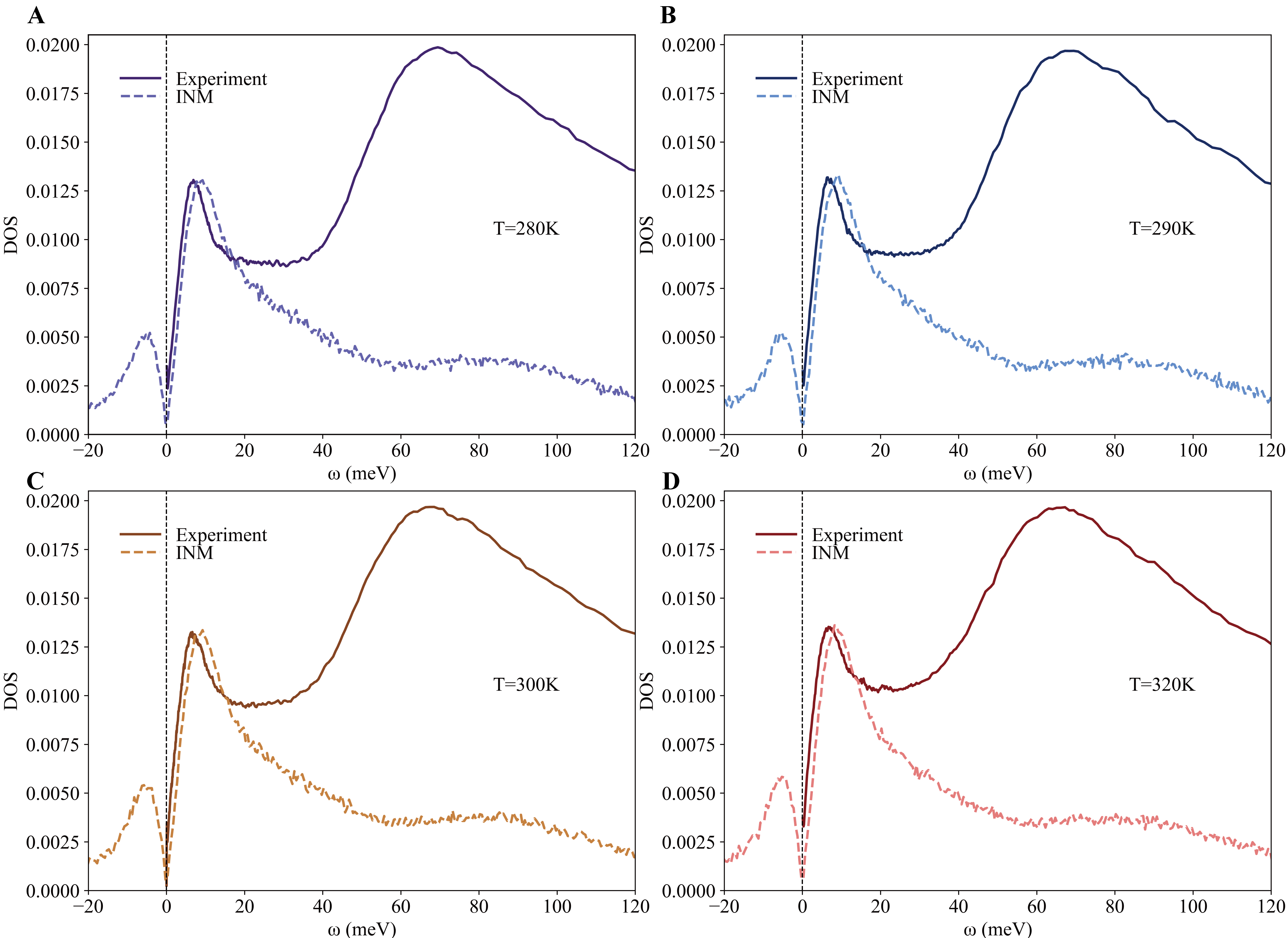}

\caption{The comparison between experimental DOS and INM DOS at \textbf{(A)} 280K (purple), \textbf{(B)} 290K (blue), \textbf{(C)} 300K (orange), \textbf{(D)} 320K (red). }
\label{fig:figure9}
\end{figure*}

In Fig.\ref{fig:figure9}, we show the comparison of the experimental DOS for water and the DOS of stable INM at different temperatures. The comparisons at $280$ K, $290$ K, $300$ K and $320$ K are shown respectively in Fig.\ref{fig:figure9}A, Fig.\ref{fig:figure9}B, Fig.\ref{fig:figure9}C and Fig.\ref{fig:figure9}D. The two DOS curves have been normalized to the first peak. The two curves are quite similar in the low frequency region. As explained before, there is a difference between the two DOS, as the INM DOS does not contain the diffusion component while the experimental DOS shows a finite value at zero frequency. As expected, the two curves are closer to each other in the low frequency region for smaller temperatures. This is simply because $g(0)$, which is not captured by the normal mode analysis, diminishes with decreasing the temperature. Above approximately $20$ meV, the two curves differ and the INM DOS shows much flatter and weaker bands, representing the stretching modes and librational motion. For all the temperatures considered, the slope of the linear low frequency regime is approximately the same in both DOS curves.

\end{document}